%% file: preprint.tex
\documentclass{llncs}
\RequirePackage{filecontents}
\PassOptionsToPackage{dvipsnames}{xcolor}
\RequirePackage{comgipp}
\RequirePackage{amsfonts}
\RequirePackage{amsmath}
\RequirePackage{authblk}

\input{07_reference.tex}

\input{header.tex}
\title{\fullTitle}

\begin{document}
\input{05_llncs_authors.tex}
\maketitle
\thispagestyle{firststyle}
  
\begin{abstract}
	\input{02_abstract.tex}

	\input{04_keywords.tex}
\end{abstract}
  
\input{03_mainmatter.tex}

\printbibliography
\end{document}

%% file: 07_reference.tex
\newcommand{\theReferenceText}{\fullcite{Trautwein2021}}

%% file: header.tex
\usepackage[
	sortcites,
	backend=biber,
	style=numeric,
	firstinits=true,
	url=false,
	isbn=false,
	safeinputenc,
]{biblatex}

\usepackage{amsmath,amssymb,amsfonts}
\usepackage{algorithmic}
\usepackage{graphicx}
\usepackage{textcomp}
\usepackage{xcolor}
\usepackage{csquotes}
\usepackage{hyperref}
\usepackage{glossaries}
\input{glossaries.tex}

\newcommand{\keywordSeparator}{, }

\input{01_title.tex}

\makeatletter
\@ifclassloaded{acmart}{%
	\title[\shortTitle]{\fullTitle}%
	}{%
	\title{\fullTitle}%
	\renewcommand{\keywordSeparator}{\and}
}
\makeatother

%% file: glossaries.tex
\newacronym{ipfs}{IPFS}{Interplanetary File-System}

\newacronym{dht}{DHT}{Distributed Hash Table}

\newacronym{mdns}{mDNS}{multicast DNS}

\newacronym{pake}{PAKE}{Password-Authenticated Key Exchange}

\newacronym{bip39}{BIP39}{39th Bitcoin Improvement Proposal}

\newacronym{id}{ID}{identifier}

\newacronym{cid}{CID}{Content Identifier}

\newacronym{nat}{NAT}{Network Address Translation}


%% file: 01_title.tex
\newcommand{\fullTitle}{%
	Introducing Peer Copy $-$ A Fully Decentralized Peer-to-Peer File Transfer Tool
}
\newcommand{\shortTitle}{%
	Introducing Peer Copy
}

%% file: 05_llncs_authors.tex
\author{Dennis Trautwein\inst{1}\orcidID{0000-0002-8567-2353} \and Moritz Schubotz\inst{2}\orcidID{0000-0001-7141-4997} \and Bela Gipp\inst{1}\orcidID{0000-0001-6522-3019}}

\authorrunning{D. Trautwein et al.}
\institute{Data and Knowledge Engineering,\\University of Wuppertal, Germany, \email{\{last\}@gipplab.org}\\ \and
	Leibniz Institute for Information Infrastructure,\\FIZ-Karlsruhe, Berlin, Germany
	\email{\{first\}.\{last\}@fiz-karlsruhe.de}}

%% file: 02_abstract.tex
\textit{Peer Copy} is a decentralized, peer-to-peer file transfer tool based on \texttt{libp2p}.
It allows any two parties that are either both on the same network or connected via the internet to transfer the contents of a file based on a particular sequence of words.
Peer discovery happens via multicast DNS if both peers are on the same network or via entries in the distributed hash table (DHT) of the InterPlanetary File-System (IPFS) if both peers are connected across network boundaries.
As soon as a connection is established, the word sequence is used as the input for a password-authenticated key exchange (PAKE) to derive a strong session key.
This session key authenticates the peers and encrypts any subsequent communication.
It is found that the decentralized approach to peer-to-peer file transfer can keep up with established centralized tools while eliminating the reliance on centralized service providers.

%% file: 04_keywords.tex
\keywords{%
	libp2p \keywordSeparator
	ipfs \keywordSeparator
	peer-to-peer \keywordSeparator
	p2p \keywordSeparator
	file transfer \keywordSeparator
	DHT
}

%% file: 03_mainmatter.tex
\section{Introduction}\label{sec:introduction}

Individual file transfer seems to be a solved problem with processes ranging from physically transporting thumb drives to protocols like \texttt{ftp}\footnote{\href{https://www.rfc-editor.org/rfc/rfc959.txt}{https://www.rfc-editor.org/rfc/rfc959.txt}} or \texttt{smtp}\footnote{\href{https://www.rfc-editor.org/rfc/rfc5321.txt}{https://www.rfc-editor.org/rfc/rfc5321.txt}}, to commercial services like Dropbox\footnote{\href{https://www.dropbox.com/ }{https://www.dropbox.com/}}.
Yet all of these tools require an inconvenient setup procedure.
You need to be in physical possession of the thumb drive, you need to have the destination server properly configured to accept \texttt{ftp} or \texttt{smtp} requests, or both peers need to have an account at the same file hosting service.
A set of tools, most notably \texttt{croc}\footnote{\href{https://github.com/schollz/croc}{https://github.com/schollz/croc}} and \texttt{magic-wormhole}\footnote{\href{https://github.com/magic-wormhole/magic-wormhole}{https://github.com/magic-wormhole/magic-wormhole}}, solve this problem by only requiring the user to transmit a short passphrase to the receiving peer to initiate a file transfer.
They allow the transmission of data without special knowledge about the technical infrastructure employed by the peers.
These tools, however, rely on a small set of servers which are usually operated by the maintainers of the open-source projects to orchestrate peer discovery and data relaying~\cite{BrianYoutube16}.
This model of operation poses centralization concerns and puts the service's sustainable operation in question as a recent issue in the \texttt{croc} repository shows\footnote{\href{https://github.com/schollz/croc/issues/289}{https://github.com/schollz/croc/issues/289}}.
The small set of private servers constitute single points of failure and an attack target to disrupt the service.
Further, the service operators have the power over whom to serve and can gather extensive knowledge about communication patterns.

In this paper, we present \textit{Peer Copy} (\texttt{pcp}) -- a decentralized, peer-to-peer file transfer tool based on \texttt{libp2p}\footnote{\href{https://libp2p.io/}{https://libp2p.io/}}.
Many concepts like the command-line user interface and user experience, as well as the \gls{pake}~\cite{Boneh2020} and the concept of channels (explained later in section \ref{sec:functionality}) were adapted or reused from \texttt{croc} and \texttt{magic-wormhole}.
The novelties of this tool are the extensive architectural differences in the peer discovery and data relaying mechanisms that render centralized server infrastructure obsolete.
The main contribution of this paper is \textbf{a decentralized peer discovery mechanism based on low entropy passphrases}.

During usual operation, the \texttt{pcp} process lifecycle can be separated into the stages of peer discovery, peer authentication, and file transfer.
The novelty of \texttt{pcp} lies in the decentralized peer discovery mechanism, which employs \gls{mdns}\footnote{\href{https://www.rfc-editor.org/rfc/rfc6762.txt}{https://www.rfc-editor.org/rfc/rfc6762.txt}} and the \gls{dht} from the \gls{ipfs}~\cite{Benet14}.
Peer authentication is done via \gls{pake}, where a small number of random words, e.g., four, from the \gls{bip39}\footnote{\href{https://github.com/bitcoin/bips/blob/master/bip-0039.mediawiki}{https://github.com/bitcoin/bips/blob/master/bip-0039.mediawiki}} serve as a passphrase.
File transfer can either be direct or transitive through a \texttt{libp2p} relay node.

The following section \ref{sec:functionality} describes the functionality and explains how each aforementioned concept relates to and facilitates the file transfer capabilities of \texttt{pcp}.
Section \ref{sec:conclusion} gives an outlook and future improvement opportunities.

\section{Functionality}\label{sec:functionality}

\texttt{pcp} provides two top-level commands \texttt{pcp send} and \texttt{pcp receive}.
When running \texttt{pcp send}, the user is presented with four words that need to be transferred, e.g., spoken or digitally, to a peer who will then use these words to run \texttt{pcp receive the-four-random-words}.
Both instances of the \texttt{pcp} process initiate the discovery procedure and try to find each other.
After successful connection, authentication, and manual confirmation, the file gets transferred.

Peer discovery, among relay services, is one of the mechanisms that poses centralization concerns in established tools like \texttt{croc} and \texttt{magic-wormhole}.
\texttt{pcp} employs \gls{mdns} and a novel \gls{dht}-based approach to enable the discovery of the desired peer.
In the following section, we focus on the \gls{dht}-based approach and later extend the discussion to \gls{mdns}.

\paragraph{Discovery}\label{sec:functionality.discovery}

Peer discovery works by devising an identifier that can be constructed by both parties solely based on shared information and information that can be derived independently. For the former, \texttt{pcp} uses the aforementioned word sequence, and for the latter, the current system time. This identifier is then used as a rendez-vous point in the \gls{dht}.

\begin{figure}
	\centering
	\includegraphics[width=.8\linewidth]{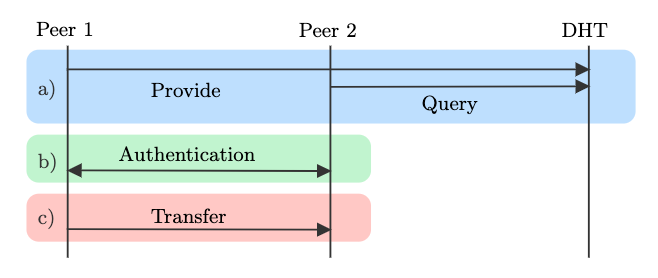}
	\caption{Different stages of the \texttt{pcp} file transfer protocol. In the depicted scenario \textit{Peer 1} wants to transfer a file to \textit{Peer 2}.}
	\label{fig:stages}
\end{figure}

When the user states the intention to initiate a file transfer by running \texttt{pcp send FILE}, four words from the \gls{bip39} English wordlist are chosen at random.
The words are claimed to be easily memorable, typeable, and pronounceable.
The first word is interpreted as a numeric channel \gls{id} by taking its index in the wordlist in the range of $0$ to $2047$.
\texttt{pcp} uses the channel \gls{id} and the current unix system timestamp to generate the discovery \gls{id}: \texttt{/pcp/\{timestamp\}/\{channel-identifier\}}.
It then puts the \gls{cid}\footnote{\url{https://docs.ipfs.io/concepts/content-addressing/}} of that discovery \gls{id} as a \enquote{provider record} in the DHT, indicating that it possesses the associated data and is willing to provide it (Fig. \ref{fig:stages}a).
The timestamp is used to limit collisions of the discovery \gls{id} between multiple simultaneously and independently operating peer pairs as provider records can stay in the \gls{dht} for up to 24 hours.
Further, it is truncated to the most recent $5$-minutes time slot to adjust for clocks that are out-of-sync, word sequence transmission times and user latency.
Note that provider records usually contain the \glspl{cid} of the underlying data, which has integrity validation advantages.
This is different in the case of \texttt{pcp} because the string is not based on the content to be transmitted.
It is solely used for discovery purposes.


On the receiving side, the user runs \texttt{pcp receive the-four-random-words}, constructs the identical discovery \gls{id}, and queries the \gls{dht} for providers of the associated \gls{cid} (Fig. \ref{fig:stages}a).
To further mitigate clock synchronization issues and time slot boundary problems, the peer also queries the \gls{dht} in parallel for the previous $5$ minute time slot.
As soon as the peer has found the \gls{dht} provider record \texttt{libp2p} handles the resolution to an associated publicly reachable IP-address. It uses the \enquote{peer records} in the \gls{dht} that were automatically populated by \texttt{libp2p} on the sending side. These peer records may contain \texttt{libp2p} nodes that have advertised relaying capabilities\footnote{\href{https://docs.libp2p.io/concepts/circuit-relay/}{https://docs.libp2p.io/concepts/circuit-relay/}}. This addresses the second centralization concern elaborated above.

When users want to transmit files on the local network \texttt{pcp} employs \gls{mdns} for peer discovery.
This is completely transparent to the user as this discovery is started in parallel to the \gls{dht}-based mechanism.
\texttt{pcp} uses \gls{mdns} for advertising the discovery \gls{id} from above in the current network.
The receiving peer issues \gls{mdns} queries for the same discovery \gls{id} to find the peer on the local network.
If the query resolves, the receiving peer immediately establishes a connection and continues with the authentication steps.

\paragraph{Authentication}\label{sec:functionality.authentication}

It is still possible that two or more pairs of peers are simultaneously trying to transmit a file with the same channel \gls{id}.
This means that their discovery \glspl{id} collided and therefore may establish connections to wrong peers.
\texttt{pcp} ensures with an authentication step, based on the remaining three words, that the connection is established between the correct peers.

This authentication step is adapted from \texttt{croc} and \texttt{magic-wormhole}.
The peers use the complete list of words as the input passphrase to the \gls{pake} protocol to derive a strong session key.
The peers use this session key to send each other challenges to ensure that both parties arrived at the same key  (Fig. \ref{fig:stages}b).
If the challenge failed, the connection is dropped; otherwise, the peers proceed to the file transfer stage.
Every subsequent communication is now encrypted with the strong session key.



\paragraph{Transfer}

In the file transfer stage, the sending peer first transmits information about the file to be sent to the receiving peer.
The user is prompted these information like filename and file size to confirm the file transfer.
Upon confirmation, the transfer starts  (Fig. \ref{fig:stages}c).




\section{Conclusion \& Future Work}\label{sec:conclusion}

We found that the discovery mechanism of \texttt{pcp} presents a viable alternative to established centralized methods.
Yet, \gls{dht} query resolution can lie in the order of minutes while the established tools operate in the seconds or even sub-second range.
Also, \texttt{libp2p} does not currently provide robust default mechanisms to deal with peers behind NATs which is highly relevant for \texttt{pcp}.
In the future, we plan to decrease the reliance on bootstrap nodes by, e.g., by employing a locally installed \gls{ipfs} node and conduct performance evaluations.